\def\c{\chi}
\def\d{\delta}

\def\f{\phi}

\def\l{\lambda}

\def\n{\nu}
\def\o{\omega}
\def\p{\pi}
\def\r{\rho}

\def\beg{\begin{equation}}
\def\eeq{\end{equation}}
\def\begr{\begin{eqnarray}}
\def\eeqr{\end{eqnarray}}

\documentstyle[12pt]{article}
\begin{document}
\begin{center}
{\Large{\bf Multiple gaps in quantum Hall effect including
review of prl 1998-2001 papers}}
\vskip0.35cm
{\bf Keshav N. Shrivastava}
\vskip0.35cm
{\it School of Physics, University of
Hyderabad, Hyderabad 500046, India}
\end{center}
\begin{abstract}
The fractions of constant conductivity, $i$, where the
conductivity is $ie^2/h$ are interpretted to arise from the
summation of two frequencies, $\o_1+\o_2$, type of
processes so that the quantum Hall effect becomes a problem of
spectroscopic energy levels. We describe the conversion of
resistivity to the frequency spectrum. An effort is made to look
at the recent articles from the Phys.Rev.Lett. 1998-2001 to see
if there is any data which does not agree with our theory. The
theoretical wave functions which claim to provide the 
theoretical base are found to be irrelevant to the 
experimental data.
\end{abstract}
\newpage
\baselineskip22pt
\section{Introduction.}
Recently, Eisenstein et al [1] have observed the quantum Hall
conductivity quantized at $\n=$7/2 and 5/2. On high field side
of 7.2 the fraction 10/3 is cleanly visible. Similarly, 11/3 is
seen at low field side. As we have explained 10/3 and 11/3 are
particle-hole Kramers conjugates [2]. The fractions 16/5 and
19/5 are also clearly visible which require a suitable
explanation. It has been noted that there is a spin branching.
We make an effort to understand these fractions and the spin
branching. Some of the data is obtained by illuminating the
sample with a red light emitting diode. After a finite time, the
diode is switched off and then the resistivity is recorded in
the Hall geometry. The effect of light is therefore to populate
some of the excited states which affect the resistivity.

In this paper, we interpret the fractions 7/2, 10/3, 11/3, 16/5
and 19/5 and their particle-hole symmetry and hence the spin
branching. We also look at 1998-2001 issues of the Phys. Rev.
Lett. to see if there are any contractions between our theory
and the data. It is found that in all of the cases our theory is
in agreement with the data.

\section{\bf Summation process.}
In our theory the conductivity is quantized in fractions
determined by two series [3]. One of the series is the same as
half of the Lande's g value and the other series is obtained by
reversing the sign of the spin in the Lande's formula.
Therefore, the theory is sufficiently simple, free of errors and
exact. The resistivity is obtained from the spectrum. Therefore,
the plateau width is related to the line width in the spectrum
which arises from the many-body interactions. The half of the
Lande's g-value gives one series,
\begin{equation}
\nu_+ = {l+1\over 21+1}
\end{equation}
and the other series obtained by using negative sign for the spin
is 
\begin{equation}
\nu_- = {l\over 2l+1}.
\end{equation}
Usually for conduction electrons only $l=0$ is taken. However we
have to take many finite values of $l$ so that,
\vskip0.25cm
\begin{center}
\begin{tabular}{c||cccccc}
\hline\hline
$l$ & 0 & 1 & 2 & 3 & 4 & $\infty$\\
$\n_+$ &   1 & 2/3 & 3/5 & 4/7 & 5/9 & 1/2\\
$\n_-$ &   0 & 1/3 & 2/5 & 3/7 & 4/9 & 1/2\\
\hline\hline
\end{tabular}
\end{center}
\vskip0.2cm
\noindent predicts all of the fractions. As the fractions give energies,
$n\nu_{\pm}$ are also possible, where $n$ is an integer. We can
identify $n$ as the Landau level quantum number,$n$. The values
$\nu_+$ and $\nu_-$ are Kramer's congugate states. Multiplying
the above numbers by $n$ gives all of the integers when we
multiply 1 by $n$ which is the case for $\nu_+$ with $l=0$. This
is of course the usual conduction band which gives the quantum
Hall step at 1, for which the resistivity is $h/e^2$. The
$l\to\infty$ limit of the two series is 1/2. When we multiply
this value by $n$ we get $n/2$. This value is symmetric in the
sense that one series is on the high-field side of $n/2$ and the
other series is on the left. Thus for any given value of $n$ we
have both the series and the patterns of a center with left and
right sides is repeated for all other values of $n$. All these
predicted features are the same as found experimentally in Fig.
18 of Stormer [4]. Therefore, it is believed that the theory is
correct. In fact Willett et al [5] have used our series but did
not give reference to our paper. We will now explain how the
above two series can be added to derive the experimentally
observed values. We consider $\o_1+\o_2$ type summation process
so that $\o_1+n\o_2$ type frequencies can also be observed. In
order to generate 2 in the denominator we make use of $n/2$. Now
we take $\l\to\infty$ limit and multiply it by 7 which gives
7/2. The fraction 1/3 has negative spin from above. Therefore,
we obtain, 3+(1/3)=10/3 which is observed on the right hand
side of 7/2. The positive spin gives 2/3 for $l=1$ in the $\n_+$
series and hence we predict, 3+(2/3)=11/3 which is observed on
the left hand side of 7/2. Therefore, 10/3 and 11/3 are the
Kramers particle-hole conjugate states with spin reversed. This
provides the explanation for 11/3, 7/2 and 10/3 in terms of
values as well as the symmetries. The difference between 11/3
and 7/2 and that between 7/2 and 10/3 is exactly equal so that
7/2 is in the centre of 10/3 and 11/3. The fraction 10/3 uses
1/3 which has negative spin while 11/3 uses 2/3 which has
positive spin. Therefore, we are able to identify the spin
branch correctly as displayed in Fig. 1.

Let us now look at 16/5 which may be written as 3+(1/5). For
$l=2$ for $\n_+$ we have 3/5 and for $\n_-$ we have 2/5. The
difference between 3/5 and 2/5 is 1/5. For $\n=3$ we obtain
$\o_3$, for 3/5 we have $\o_{3/5}$ and for 2/5 we have
$\o_{2/5}$. The difference $\o_{3/5}-\o_{2/5}$ generates
$\o_{1/5}$ and hence $3+{1\over5}$ is generated by
$\o_3+(\o_{3/5}-\o_{2/5})$. In this case the spin is mixed and
it is not generated from 7/2.

We consider the next fraction 19/5 at which there is a plateau in
the transverse resistivity. This fraction may be written by
taking $\n_-=2/5$ so that the spin is negative. For $l=2$, we
obtain this value which we multiply by $n=2$ which means that we
have a level at $n=2, l=2$ with spin negative so that we get
$n\times2/5=4/5$. We consider the frequency due to $\n=3$ and
add $\o_3+\o_{4/5}$ to obtain 3+4/5=19/5. In this way we are
able to assign the spin to 19/5 and also understand it in terms
of a summation process. Thus our theory explains 16/5 as well as
19/5.

\section{Angular momentum.}

The theory given in ref. 3 thus gives the same values for the
fractions as those experimentally observed by Stormer [4] and
Willett et. al [5]. It gives the spin value and hence couples
the spin with charge by use of Bohr magneton. It also describes
the Kramers particle-hole symmetry. We have compared this theory
with a lot of experimental data and found that in all cases the
fractions tabulated in ref. 3 are the same as found in the
experimental data [6]. The high Landau levels are easily
explained by this theory [7]. In fact, the theory of ref. 3
works so well in terms of getting agreement with the
experimental data that many of its features are described in a
book [8]. The interpretation of high Landau levels 5/2, 7/3 and
8/3 has been obtained and 7/3 and 8/3 are found to be
particle-hole conjugate states [9]. The relative change in
resistivity is given by [10],
\beg
{\d R_{xx}\over R_{xx}} = 4\p \c^{\prime\prime} Q
\eeq
where $\c^{\prime\prime}$ is the imaginary part of the
susceptibility and $Q$ is the quality factor. The imaginary part
of the susceptibility resonates with the relaxation rate $T_2$
and resonance frequency $\o_o$ as
\begr
\c^{\prime\prime} &=& \c^\prime-i\c^{\prime\prime}\\
\c^{\prime\prime} &=& {\c_o\over2} \o_oT_2{1\over1+(\o-\o_o)^2T^2_2}
\,\,.
\eeqr
As one goes from one level to another, the peaks of
$\c^{\prime\prime}$ appear as peaks in the resistivity. The
frequency $\o$ is obviously proportional to the magnetic field.
In fact, the table of $\n_{\pm}$ given above predicts a soft
mode of zero frequency so that we expect a Goldstone mode to
emerge when the magnetic length is equal to the separation
between two layers of a semiconductor [11]. When the sample are
first illuminated by a red diode and the resistivity is recorded
after switching off the light, considerable changes occur in the
populations of the excited states which affect the resistivity
data. In some cases, this type of experiment results into
improved resolution [12]. The theoretical work predicts that
there is a small correction to the value of the Bohr magneton
[13]. The polarization of the half-filled Landau level is also
well explained [14]. 

\section{Discussion and the list of 1998-2001 prl papers.}

We have to consider if our theory agrees with Laughlin's theory
of fractional charge. Unfortunately, the problem depends on the
flux quantization so that the field multiplied by an area is a 
constant,
\beg
B.A = {hc\over e}
\eeq
replacing $e$ by $(1/3)e$ is  exactly equivalent to replacing 
area $A$ by $(1/3)A$. Therefore, it is not necessary to
change $e$ to $(1/3)e$. This problem is hidden in Laughlin's
theory of an area operator. Since the product $eA$ occurs,
Laughlin could have chosen the (1/3) of the area rather than
(1/3) of the charge. Hence, Laughlin's theory is of no relevance
to the experimental data of quantum Hall effect [15].

There are non-Abelian theories [16-18] and topology [19] has
been applied. In the Maxwell equations apparently it is possible
to make a correction to the vector potential if a correction is
also made to the scalar potential. These corrections are called
Chern-Simon fields which shift the magnetic field [20-21]. Many
problems related to charging have been reported [22-31]. The
scaling theory have been discussed [32-40]. There are
discussions of  spin-orbit interaction [41-42]. As the
resistivity oscillates between low and high values, the high
values are associated with the terminology of an insulator and
metal-insulator transition has been discussed [43-53]. Detailed
work on conductivity [34-61] with specific fractions, integers
[62-65] as well as (1/2)-fractions [66-73] is available. A large
amount of theoretical work has been published to search for the
correct interpretation of data but there is not much success. In
one case theory of fractals [74], is applied to the curves of
the resistivity but this theory neither agrees with the data nor
is relevant in any way but the effort goes on. In any case
quantum Hall effect is not represented by fractals. In the
composite fermion (CF) theory [75-96], ``even number'' of fluxes
are attached to the electron, i.e., two flux quanta are attached
to one electron to explain the plateau in the Hall resistivity
at $\n=1/3$. The magnetic field becomes $B^*=B-2\r\f_o$ where
$\f_o=hc/e$ and $\r$ is the charge density. The correct way to
quantize the field $B$ is $BA=n\f_o$ and not by $B^*$, where $A$
is an area and $n$ an integer. When we make a coil with a
current passing through the coil, a magnetic field is produced.
This is called the Biot and Shavart's law in classical
electrodynamics as shown in Fig. 2a. If a flux is attached to
the electron it will affect the field and the Biot and Shavart's
law will be modified as shown in Fig. 2b. We also show the even
number of fluxes attached to one electron in Fig. 2c. The idea
of attaching the flux tubes to electrons was considered by
Wilczek [97]. The experimental data does not show the ``even
number'' quantization. As far as the statistics is concerned, we
find [98] that the composite fermions actually have a component
of bosons whereas the theory of ``composite fermions'' demands
that there quasiparticles be fermions. It is clear that
composite fermions do not obey the Biot and Shavart's law. Some
experimentalists have actually claimed to find agreement between
the composite fermions and the data. These experimental claims
are obviously deceptive or incorrect. The composite fermion
theory is not a physically observable theory and it is of no
relevance to the experimental data on quantum Hall effect. It
should be made clear that fluxes are not attached to the
electrons which is another way of saying that the composite
fermions are not observed.

Several workers have measured [99-106] the nuclear magnetic
resonance near the magnetic field value where plateaus occur in
the quantum Hall effect. These experiments are able to find the
polarization from the Kinght shift [14]. It is clear that the
temperature at which the paramagnetic electrons become
ferromagnetic is zero. Therefore, the ferromagnetic phase is not
found. There are several theories [107-115] of ferromagnetic
phase in the quantum Hall effect but these theoretical claims of
finding ferromagnetic phase in the quantum Hall effect are not
correct because the Curie-Weiss critical temperature is zero.
There is a considerable effort to understand edges [116-128] and
stripes [129-193]. A Goldstone mode has been found [134-135]
experimentally at half-filled Landau level for which several
theorists have constructed [136-150] their models but due to lack
of good foundation their ideas are fragmentary and hence not
applicable to the data. The correct theory of the Goldstone mode
in accordance with the experimental data, is given by
Shrivastava [11]. The half-filled Landau level presents some
analogy with the Bose-Einstein condensation [151-158] and
levitating states have been reported [154].

The spin and charge fluctuations in the fluid are subject to
exchange interaction which can lead to a ferromagnetic state.
There is a phase transition when the Lande's $g$ value
approaches zero. The quasiparticles in this limit are called
``skyrmions'' [155,156]. An effort has been made to suggest that
``skyrmions'' are important for the quantum Hall effect
[157-163]. However, the experimental evidence is not in favour
of exchange interaction which gives a Curie temperature. The
effect of spin has been considered but the theories are not in
accord with the data [164-172]. Some authors [173-177] are
arguing in favour of spin-charge decoupling in which the
fermions and bosons are exchanged so that the statistics is
mixed in one dimension but there is no experimental evidence in
favour of such a phenomenon. The low field high integer Landau
levels have been discussed and considerable effort has been made
to find the interpretation [178-181]. In this case the angular
momentum theory is in agreement with the experimental data [7].
The problem of dimensionality has been discussed. Although the
electron gas is two dimensional, the quantum Hall effect 
appears to require three dimensions [182-184]. Our theory [3]
also does not use the dimensionality explicitly. The emission of
phonons has been studied in the quantum Hall effect  [185-186]
and the relaxation is the result of phonon emission and
absorption. Light scattering experiments [187-188] gave the
usual results and evidence was found to support fractionally
charged quasiparticles. The experimental measurements of mass
and $g$ values [189-190] show that some of the masses are equal.
We have explained that this equality arises from the
particle-hole symmetry [2]. There is emphasis on the anisotropy
of the currents [191-192] but the polynomials which determine
the anisotropy have not yet been determined. The tunneling
current [193-194] magnetization [195-196] and the drag effects
have been discussed [197-198]. There are some indications of
rotations [199-200]. An accurate measurement of the Planck's
constant has been reported [201].

\section{Conclusions.}

There are a large variety of experimental measurements in the
quantum Hall effect. In all the cases our theory of angular
momentum first published in ref. 3 gives the correct
interpretation of the experimental data. The apparent fractional
charge is determined in terms of spin and orbital angular
momenta. There is no need of a wave function of a fractionally
charged quasiparticle as it is clear that fractionally charged
quasiparticles do not occur in the quantum Hall effect.

\section{References.}
\baselineskip14pt
\begin{enumerate}
\item J.P. Eisenstein, K.B. Cooper, L.N. Pfeiffer and K.W. West,
	arXiv; Cond-mat/0110477.
\item K.N. Shrivastava, Mod. Phys. Lett. {\bf13}, 1087 (1999).
\item  K.N. Shrivastava, Phys. Lett. A{\bf113}, 435 (1986);
	{\bf115}, 459(E) (1986).
\item H.L. Stormer, Rev. Mod. Phys. {\bf71}, 875 (1999).
\item R.L. Willett, K.W. West and L.N. Pfeiffer, Phys. Rev.
	Lett. {\bf75}, 2988 (1995) [This paper used the same series as
	in ref. 3].
\item K.N. Shrivastava, in Frontiers of Fundamental Physics 4,
	edited by B.G. Sidharth and M.V. Altaisky, Kluwer
	Academic/Plenum Pub. New York 2001.
\item K.N. Shrivastava, Mod. Phys. Lett. B{\bf14}, 1009 (2000).
\item K.N. Shrivastava, Superconductivity: Elementary Topics, World
	Scientific, New Jersey 2000.
\item K.N. Shrivastava, cond-mat/0103604.
\item  K.N. Shrivastava, cond-mat/0104577.
\item K.N. Shrivastava, cond-mat/0106445.
\item K.N. Shrivastava, cond-mat/0109407.
\item K.N. Shrivastava, cond-mat/0104004.
\item K.N. Shrivastava, cond-mat/0106160.
\item R.B. Laughlin, Phys. Rev. B{\bf27}, 3383 (1983).
\item E. Ardonne and K. Schoutens, Phys. Rev. Lett. {\bf82},
5096 (1999).
\item K. Schoutens, Phys. Rev. Lett. {\bf81}, 1929 (1998).
\item D.A. Ivanov, Phys. Rev. Lett. {\bf86}, 268 (2001).
\item M. Oshikawa, Phys. Rev. Lett. {\bf84}, 1535 (2000).
\item F. Faure, B. Zhilinskii, Phys. Rev. Lett. {\bf85},
	960 (2000).
\item A. Le Clair, Phys. Rev. Lett. {\bf84}, 1292 (2000).
\item D.H. Cobden, C.H.W. Barnes and C.J.B. Ford, Phys. Rev.
	Lett. {\bf82}, 4695 (1999).
\item E. Yahel, A. Tsukernick, A. Palevski and H. Shtrikman,
 Phys. Rev. Lett.
	{\bf81}, 5201 (1998).
\item J. Ye and  S. Sachdev, Phys. Rev. Lett. {\bf80}, 5409 (1998).
\item N. Freytag, Y. Tokunaga and M. Horvatic, Phys. Rev. Lett.
	{\bf87}, 136801 (2001).
\item T.G. Griffiths, E. Comforti, M. Heiblum, A. Stern
 and V. Umansky, Phys. Rev. Lett. {\bf85}, 3918 (2000).
\item J.P. Watts, A. Usher, A.J. Matthews, M. Zhu, M. Elliott,
W. G. Herrenden-Harker, P. R. Morris, M. Y. Simmons and 
D. A. Ritchie, Phys. Rev. Lett. {\bf81}, 4220 (1998).
\item G. Yusa, H. Shtrikman, I. Bar-Joseph, Phys. Rev. Lett.
	{\bf97}, 216402 (2001).
\item M. Kataoka, C.J.B. Ford, G. Faini,D. Mailly, M. Y. Simmons, D. R. Mace, C. T. Liang and D. A. Ritchie, Phys. Rev. Lett.
	{\bf83}, 160 (1999).
\item T.H. Oosterkamp, J.W. Janssen, L.P. Kouwenhoven, D. G. Austing, T. Honda and S. Tarucha,  Phys. Rev. Lett. {\bf82}, 2931 (1999).
\item L. Gravier, M. Potemski, P. Hawrylak, and B. Etienne, Phys. Rev. Lett.
	{\bf80}, 3344 (1998).
\item J.J. Mares, J. Kristofik and P. Hubik, Phys. Rev. Lett.
	{\bf82}, 4699 (1999).
\item I.A. Gruzberg, A.W.W. Ludwig and N. Read, Phys. Rev. Lett.
	{\bf82}, 4524 (1999).
\item N.Q. Balaban, U. Meirav, I. Bar-Joseph, Phys. Rev. Lett.
	{\bf81}, 4967 (1998).
\item D.N. Sheng and Z.Y. Weng, Phys. Rev. Lett. {\bf80}, 580 (1998).
\item J.E. Moore, A. Zee and J. Sinova, Phys. Rev. Lett.
	{\bf87}, 046801 (2001).
\item J.B. Marston and  S.W.T. Sai, Phys. Rev. Lett. {\bf82}, 4906 (1999).
\item S. Kettemann and  A. Tsvelik, Phys. Rev. Lett. {\bf82}, 3689 (1999).
\item M.A. Eriksson, A. Pinczuk, B.S. Dennis, Phys. Rev. Lett.
	{\bf82}, 2163 (1999).
\item D.N. Sheng, Z.Y. Weng, Phys. Rev. Lett. {\bf83}, 144 (1999).
\item V.I. Fal'ko and S.V. Iordanskii, Phys. Rev. Lett. {\bf84},
	127 (2000).
\item E.N. Bulgakov, K.N. Pichugin, A.F. Sadreev, Phys. Rev.
	Lett. {\bf83}, 376 (1999).
\item G. Xiong, S.D. Wang, Q. Niu, Phys. Rev. Lett. {\bf87},
	216802 (2001).
\item R.T.F. Van Schaijk, A. de Visser, S.M. Olsthoorn, Phys.
	Rev. Lett. {\bf84}, 1567 (2000).
\item R.J. Nicholas, K. Takashina, M. Lakrimi, Phys. Rev. Lett.
	{\bf85}, 2364 (2000).
\item X-G. Wen, Phys. Rev. Lett. {\bf84}, 3950 (2000).
\item B. Huckestein, M. Backhaus, Phys. Rev. Lett. {\bf82}, 5100
	(1999).
\item L.P. Pryadko, A. Auerbach, Phys. Rev. Lett. {\bf82}, 1253 (1999).
\item L.B Rigal, D.K. Maude, M. Potemski, Phys. Rev. Lett. {\bf82},
	1249 (1999).
\item D.G. Polyakov, K.V. Samokhin, Phys. Rev. Lett. {\bf80},
	1509 (1998).
\item V. Kagalovsky, B. Horovitz, Y. Avishai, Phys. Rev. Lett.
	{\bf82}, 3516 (1999).
\item S.C. Dultz, H.W. Jiang, Phys. Rev. Lett. {\bf84}, 4689 (2000).
\item M.Y. Simmons, A.R. Hamilton and M. Pepper, Phys. Rev.
	Lett. {\bf84} 2489 (2000).
\item F. Hohls, U. Zeitler, R.J. Haug, Phys. Rev. Lett. {\bf86},
	5124 (2001).
\item I. Safi, P. Devillard, T. Martin, Phys. Rev. Lett.
	{\bf86}, 4628 (2001).
\item A.M.C. Valkering, P.K.H. Sommerfeld, R.A.M. Van de Ven,
	Phys. Rev. Lett. {\bf81}, 5398 (1998).
\item B. Jovanovic and Z. Wang, Phys. Rev. Lett. {\bf81}, 2767 (1998).
\item H. Yi, H.A. Fertig, R. Cote, Phys. Rev. Lett. {\bf85},
	4156 (2000).
\item S.B. Isakov, T. Martin, S. Ouvry, Phys. Rev. Lett.
	{\bf83}, 580 (1999).
\item E. Shimshoni, A. Auerback, A. Kapitulnik, Phys. Rev. Lett.
	{\bf80}, 3352 (1998).
\item Y. Hatsugai, K. Ishibashi, Y. Morita, Phys. Rev. Lett.
	{\bf83}, 2246 (1999).
\item G. Murthy, Phys. Rev. Lett. {\bf85}, 1954 (2000).
\item M. Onoda, T. Mizusaki, T. Otsuka, Phys. Rev. Lett.
	{\bf84}, 3942 (2000).
\item B. Paredes, C. Tejedor, L. Brey, Phys. Rev. Lett. {\bf83},
	2250 (1999).
\item Y-Q. Song, B.M. Goodson, K. Maranowski, Phys. Rev. Lett.
	{\bf82}, 2768 (1999).
\item X.Y. Lee, H.W. Jiang, W.J. Schaff, Phys. Rev. Lett.
	{\bf83}, 3701 (1999).
\item T. Morinari, Phys. Rev. Lett. {\bf81}, 3741 (1998).
\item W. Pan, J-S. Xia, V. Shvarts, Phys. Rev. Lett. {\bf83},
	3530 (1999).
\item K. Ishikawa, N. Maeda, T. Ochiai, Phys. Rev. Lett.
	{\bf82}, 4292 (1999).
\item R.H. Morf, Phys. Rev. Lett. {\bf80}, 1505 (1998).
\item R. Shankar, Phys. Rev. Lett. {\bf84}, 3946 (2000).
\item W. Pan, R.R. Du, H.L. Stormer, Phys. Rev. Lett. {\bf83},
	820 (1999).
\item D.-H. Lee, Phys. Rev. Lett. {\bf80}, 4745 (1998).
\item C. Albrecht, J.H. Smet, K. Von Klitzing, Phys. Rev. Lett.
	{\bf86}, 147 (2001).
\item J.K. Jain, Phys. Rev. Lett. {\bf63}, 199 (1989).
\item K. Park and J.K. Jain, Phys. Rev. Lett. {\bf81}, 4200 (1998).
\item R.K. Kamilla, X.G. Wu and J.K. Jain, Phys. Rev. Lett.
	{\bf76}, 1332 (1996).
\item K. Park and J.K. Jain, Phys. Rev. Lett. {\bf80}, 4237 (1998).
\item J.K. Jain and R.K. Kamila, Int. J. Mod. Phys.{\bf11}, 2621 (1997).
\item K. Park and J.K. Jain, Phys. Rev. Lett. {\bf84}, 5576 (2000).
\item G. Murthy, Phys. Rev. Lett. {\bf84}, 350 (2000).
\item H.-S. Sim, K.J. Chang, G. Ihm, Phys. Rev. Lett. {\bf82},
596 (1999).
\item S. Curnoe, P.C.E. Stamp, Phys. Rev. Lett. {\bf80}, 3312 (1998).
\item A.D. Mirlin, D.E. Polyakov, P. Wolfle, Phys. Rev. Lett.
	{\bf80}, 2429 (1998).
\item F.V. Oppen, A. Stern, B.I. Halperin, {\bf80}, 4494 (1998).
\item N.K. Wilkin, J.M.F. Gunn, Phys. Rev. Lett. {\bf84}, 6 (2000).
\item I.V. Kukushkin, K.v. Klitzing, K. Eberl, Phys. Rev. Lett.
	{\bf82}, 3665 (1999).
\item N.E. Bonesteel, Phys. Rev. Lett. {\bf82}, 984 (1999).
\item J.H. Smet, K.v. Klitzing, D. Weiss, Phys. Rev. Lett.
	{\bf80}, 4538 (1998).
\item S. Zelakiewicz, H. Noh, T.J. Gramila, Phys. Rev. Lett.
{\bf85}, 1942 (2000).
\item S.D.M. Zwerschke, R.R. Gerhardts, Phys. Rev. Lett.
{\bf83}, 2616 (1999).
\item J.H. Smet, S. Jobst, K.v. Klitzing, Phys. Rev. Lett.
{\bf83}, 2620 (1999).
\item R. Shankar, Phys. Rev. Lett. {\bf83}, 2382 (1999).
\item K. Park, N. Meskini, J.K. Jain, Phys. Rev. Lett. {\bf83},
1486 (1999).
\item R.H. Morf, Phys. Rev. Lett. {\bf83}, 1485 (1999).
\item I. Ussishkin, A. Stern, Phys. Rev. Lett. {\bf81}, 3932 (1998).
\item F. Wilczek, Phys. Rev. Lett. {\bf49}, 957 (1982).
\item K.N. Shrivastava, cond-mat/0105559 (29 May 2001).
\item A.M. Song, P. Omling, Phys. Rev. Lett. {\bf84}, 3145 (2000).
\item I.V. Kukushkin, J.H. Smet, K. von Klitzing, Phys. Rev.
Lett. {\bf85}, 3688 (2000).
\item P. Khandelwal, N.N. Kuzma, S.E. Barrett, Phys. Rev. Lett.
{\bf81}, 673 (1998).
\item S. Kronmuller, W. Dietsche, K. von Klitzing, Phys. Rev.
Lett. {\bf82}, 4070 (1999).
\item S. Kronmuller, W. Dietsche, J. Weiss, Phys. Rev.
Lett. {\bf81}, 2526 (1998).
\item V.T. Dolgopolov, A.A. Shaskin, J.M. Broto, Phys. Rev.
Lett. {\bf86}, 5566 (2001).
\item S. Melinte, N. Freytag, M. Horvatic, Phys. Rev. Lett.
{\bf84}, 354 (2000).
\item A.E. Dementyev, N.N. Kuzma, P. Khandelwal, Phys. Rev.
Lett. {\bf83}, 5074 (1999).
\item A.G. Green, Phys. Rev. Lett. {\bf82}, 5104 (1999).
\item T. Jungwirth, S.P. Shukla, L. Smrcka, Phys. Rev. Lett.
{\bf81}, 2328 (1998).
\item K. Muraki, T. Saku, Y. Hirayama, Phys. Rev. Lett. {\bf87},
196801 (2001).
\item J.H. Smet, R.A. Deutschmann, W. Wegscheider, Phys. Rev.
Lett. {\bf86}, 2412 (2001).
\item V.I. Fal'ko, S.V. Iordanskii, Phys. Rev. Lett. {\bf82},
402 (1999).
\item N.R. Cooper, Phys. Rev. Lett. {\bf80}, 4554 (1998).
\item T. Jungwirth, A.H. Mac Donald, Phys. Rev. Lett. {\bf87},
216801 (2001).
\item H.B. Chan, R. C. Ashoori, L.N. Pfeiffer, Phys. Rev. Lett.
{\bf83}, 3258 (1999).
\item F.G. Monzon, H.X. Tang, M.L. Roukes, Phys. Rev. Lett.
{\bf84}, 5022 (2000).
\item K. Ino, Phys. Rev. Lett. {\bf81}, 5908 (1998).
\item S. Takaoka, K. Oto, S. Uno, Phys. Rev. Lett. {\bf81}, 4700
(1998).
\item M. Grayson, D.C. Tsui, L.N. Pfeiffer, Phys. Rev. Lett.
{\bf86}, 2645 (2001).
\item H.-C. Kao, C.-H. Chang, X.-G. Wen, Phys. Rev Lett. {\bf83},
5563 (1999).
\item U. Zulicke, Phys. Rev. Lett. {\bf83}, 5330 (1999).
\item Y.Y. Wei, J. Weis, K.v. Klitzing, Phys. Rev. Lett.
{\bf84}, 1674 (1998).
\item K. Ino, Phys. Rev. Lett. {\bf81}, 1078 (1998).
\item L.P. Pryadko, K. Chaltikian, Phys. Rev. Lett. {\bf80}, 584
(1998).
\item A.V. Shytov, L.S. Levitov, B.I. Halperin, Phys. Rev. Lett.
{\bf80}, 141 (1998).
\item V.J. Goldman, E.V. Tsiper, Phys. Rev. Lett. {\bf86}, 5841 (2001).
\item K. Sengupta, H.-J. Kwon, V.M. Yakovenko, Phys. Rev. Lett.
{\bf86}, 1094 (2001).
\item C. Wexler, A.T. Dorsey, Phys. Rev. Lett. {\bf82}, 620 (1999).
\item O.G. Balev, P. Vasilopoulos, Phys. Rev. Lett. {\bf81},
1481 (1998).
\item E.H. Rezayi, F.D.M. Haldane, Phys. Rev. Lett. {\bf84},
4685 (2000).
\item F.v. Oppen, B.I. Halperin, A. Stern, Phys. Rev. Lett.
{\bf84}, 2937 (2000).
\item H.A. Fertig, Phys. Rev. Lett. {\bf82}, 3693 (1999).
\item A. Auerbach, Phys. Rev. Lett. {\bf80}, 817 (1998).
\item S.-Y. Lee, V.W. Scarola, J.K. Jain, Phys. Rev. Lett.
{\bf87}, 256803 (2001).
\item I.B. Spielman, J.P. Eisenstein, L.N. Pfeiffer and K. W. West, Phys. Rev.
Lett. {\bf87}, 036803 (2001).
\item I.B. Spielman, J.P. Eisenstein, L.N. Pfeiffer and K. W. West, Phys. Rev. Lett. {\bf84}, 5808 (2000).
\item E. Demler, C. Nayak, S.D. Sarma, Phys. Rev. Lett. {\bf86},
1853 (2001).
\item L. Balants, L. Radzihavsky, Phys. Rev. Lett. {\bf86}, 1825
(2001).
\item A. Sawada, Z.F. Ezawa, H. Ohno, Phys. Rev. Lett. {\bf80},
4534 (1998).
\item L. Radzihavsky, Phys. Rev. Lett. {\bf87}, 236802 (2001).
\item Y.N. Joglekar, A.H. Mac Donald, Phys. Rev. Lett. {\bf87},
196802 (2001).
\item K. Yang, Phys. Rev. Lett. {\bf87}, 056802 (2001).
\item A. Stern. S.M. Girvin, A.H. MacDonald, Phys. Rev. Lett.
{\bf86}, 1829 (2001).
\item J. Schliemann, S.M. Girvin, A.H. Mac Donald, Phys. Rev.
Lett. {\bf86}, 1849 (2001).
\item M.M. Fogler, F. Wilczek, Phys. Rev. Lett. {\bf86}, 1833 (2001).
\item J. Schliemann, A.H. MacDonald, Phys. Rev. Lett. {\bf84},
4437 (2000).
\item J. Kyriakidis, D. Loss, A.H. MacDonald, Phys. Rev. Lett.
{\bf83}, 1411 (1999).
\item L. Brey, E. Demler, S.D. Sarma, Phys. Rev. Lett. {\bf83},
168 (1999).
\item E. Demler, S.D. Sarma, Phys. Rev. Lett. {\bf82}, 3895 (1999).
\item S.P. Shukla, Y.W. Suen, M. Shagegan, Phys. Rev. Lett.
{\bf81}, 693 (1998).
\item M. Troyer, S. Sachdev. Phys. Rev. Lett. {\bf81}, 5418 (1998).
\item T.L. Ho. Phys. Rev. Lett. {\bf87}, 060403 (2001).
\item B. Paredes, P. Fedichev and J.I. Cirac, Phys. Rev. Lett.
{\bf87}, 010402 (2001).
\item N.R. Cooper, N.K. Wilkin, J.M.F. Gunn, Phys. Rev. Lett.
{\bf87}, 120405 (2001).
\item Th. Koschny, H. Potempa, L. Schweitzer, Phys. Rev. Lett.
{\bf86}, 3863 (2001).
\item T.H.R. Skyrme, Proc. Roy. Soc. A {\bf226}, 521 (1954).
\item S.L. Sondhi, A. Karlhede, S.A. Kivelson and E.H. Rezayi,
Phys. Rev. B {\bf47}, 16419 (1993).
\item P. Khandelwal, A.E. Dementyev, N.N. Kuzma, Phys. Rev.
Lett. {\bf86}, 5353 (2001).
\item J.D. Naud, L.P. Pryadko, S.L. Sondhi, Phys. Rev. Lett.
{\bf85}, 5408 (2000).
\item K. Moon, K. Mullen, Phys. Rev. Lett. {\bf84}, 975 (2000).
\item Z.F. Ezawa, Phys. Rev. Lett. {\bf82}, 3512 (1999).
\item A.J. Nederveen, Y.V. Nazarov, Phys. Rev. Lett. {\bf82},
406 (1999).
\item Yu.V. Nazarov, A.K. Khaetskii, Phys. Rev. Lett. {\bf80},
576 (1998).
\item S. Melinte, E. Grivei, V. Bayot, Phys. Rev. Lett. {\bf82},
2764 (1999).
\item K. Ino, Phys. Rev. Lett. {\bf83}, 3526 (1999).
\item K. Ino, Phys. Rev. Lett. {\bf86}, 882 (2001).
\item H. Chao, J.B. Yang, W. Kang, Phys. Rev. Lett. {\bf81},
2522 (1998).
\item N. Dupuis, V.M. Yakovenko, Phys. Rev. Lett. {\bf80}, 3618 (1998).
\item J. Cardy, Phys. Rev. Lett. {\bf84}, 3507 (2000).
\item A.H. MacDonald, Phys. Rev. Lett. {\bf83}, 3262 (1999).
\item W. Apel, Yu. A. Bychkov, Phys. Rev. Lett. {\bf82}, 3324 (1999).
\item A.V. Khaetskii, Phys. Rev. Lett. {\bf87}, 049701 (2001).
\item B.W. Alphenaar, H.O. Muller, K. Tsukagoshi, Phys. Rev.
Lett. {\bf81}, 5628 (1998).
\item A.M. Chang, M.K. Wu, C.C. Chi, Phys. Rev. Lett. {\bf86},
143 (2001).
\item M. Grayson, D.C. Tsui, L.N. Pfeifter, Phys. Rev. Lett.
{\bf80}, 1062 (1998).
\item M. Hilke, D.C. Tsui, M. Grayson, Phys. Rev. Lett. {\bf87},
186806 (2001).
\item K. Flensberg, Phys. Rev. Lett. {\bf81}, 184 (1998).
\item P. Sharma, C. Chamon, Phys. Rev. Lett. {\bf87}, 096401 (2001).
\item B. Huckestein, Phys. Rev. Lett. {\bf84}, 3141 (2000).
\item T.D. Stanescu, I. Martin, P. Phillips, Phys. Rev. Lett.
{\bf84}, 1288 (2000).
\item N. Shibata, D. Yoshioka, Phys. Rev. Lett. {\bf86}, 5755 (2001).
\item F.D.M. Haldane, E.H. Rezayi, K. Yang, Phys. Rev. Lett.
{\bf85}, 5396 (2000).
\item S.S. Murzin, A.G.M. Jansen, P.v.d. Linden, Phys. Rev.
Lett. {\bf80}, 2681 (1998).
\item D. Haude, M. Morgenstern, I. Meinel, Phys. Rev. Lett.
{\bf86}, 1582 (2001).
\item M. Koshino, H. Aoki, K. Kuroki, Phys. Rev. Lett. {\bf86},
1062 (2001).
\item S. Roshko, W. Dietsche, L. J. Challis, Phys. Rev. Lett. {\bf80},
3835 (1998).
\item L. Balicas, Phys. Rev. Lett. {\bf80}, 1960 (1998).
\item M. Kang, A. Pinczuk, B.S. Dennis, Phys. Rev. Lett.
{\bf84}, 546 (2000).
\item J.E. Avron, A. Elgart, G.M. Graf, Phys. Rev. Lett.
{\bf87}, 236601 (2001).
\item A.S. Yeh, H.L. Stormer, D.C. Tsui, Phys. Rev. Lett.
{\bf82}, 592 (1999).
\item D. Yoshioka, Phys. Rev. Lett. {\bf83}, 886 (1999).
\item W. Pan, T. Jungwirth, H.L. Stormer, Phys. Rev. Lett.
{\bf85}, 3257 (2000).
\item M.P.Lilly, K.B. Cooper, J.P. Eisenstein, Phys. Rev. Lett.
{\bf83}, 824 (1999).
\item Z. Wang, S. Xiong, Phys. Rev. Lett. {\bf83}, 828 (1999).
\item I.J. Maasilta, V.J. Goldman, Phys. Rev. Lett. {\bf84},
1776 (2000).
\item I. Meinel, T. Hengstmann, D. Grundler, Phys. Rev. Lett.
{\bf82}, 819 (1999).
\item L. Brey, Phys. Rev. Lett. {\bf81}, 4692 (1998).
\item X.G. Feng, S. Zelakiewicz, H. Noh, Phys. Rev. Lett.
{\bf81}, 3219 (1998).
\item F.v. Oppen, S.H. Simon, A. Stern, Phys. Rev. Lett.
{\bf87}, 106803 (2001).
\item M. Kang, A. Pinczuk, B.S. Dennis, Phys. Rev. Lett.
{\bf86}, 2637 (2001).
\item U. Zeitler, A.M. Devitt, J.E. Digby, Phys. Rev. Lett.
{\bf82}, 5333 (1999).
\item E.R. Williams, R.L. Steiner, D.B. Newell, Phys. Rev. Lett.
{\bf81}, 2404 (1998).
\end{enumerate}
\vskip1.0cm
\noindent{\bf Figure Captions}
\begin{description}
\item[Fig.1:] The transverse resistivity data as a function of
magnetic field showing 7/2 with 10/3 with negative spin and 11/3
with positive spin.
\item [Fig.2:] (a) The correct Biot and Shavert's law, (b) one
flux attached to one electron and (c) two fluxes attached to one
electron. The cases (b) and (c) are the modifications called the
``composite fermions'' (CF). It is obvious and (b) and (c) are
not consistent with (a).
\end{description}

\end{document}